\newcommand{\epem}{\mbox{$\mathrm{e^+e^-}$}}

\newcommand{\Ebeam}{\mbox{$E_{\mathrm{beam}}$}}

\newcommand{\ycut}{\mbox{$y_{\mathrm{cut}}$}}
\newcommand{\MW}{\mbox{$M_{\mathrm{W}}$}}
\newcommand{\MZ}{\mbox{$M_{\mathrm{Z}}$}}

\newcommand{\PZ}{\mbox{$\mathrm{Z}$}}

\newcommand{\qq}{\mbox{$\mathrm{q\overline{q}}$}}
\newcommand{\mumu}{\mbox{$\mu^+\mu^-$}}
\newcommand{\tautau}{\mbox{$\tau^+\tau^-$}}

\newcommand{\Opal}{\mbox{OPAL}}

\newcommand{\LEP}{\mbox{LEP}}
\newcommand{\ISR}{\mbox{{\sc ISR}}}

\newcommand{\NMR}{\mbox{NMR}}
\newcommand{\Leptwo}{\mbox{LEP~II}}
\newcommand{\Lepone}{\mbox{LEP~I}}
\newcommand{\KK} {\mbox{KK2f}}
\newcommand{\CEEX}{\mbox{CEEX}}
\newcommand{\Herwig}{\mbox{HERWIG}}
\newcommand{\Pythia}{\mbox{PYTHIA}}
\newcommand{\Koralw}{\mbox{KORALW}}
\newcommand{\Koralz}{\mbox{KORALZ}}
\newcommand{\Phojet}{\mbox{PHOJET}}
\newcommand{\Twogen}{\mbox{TWOGEN}}
\newcommand{\Bhwide}{\mbox{BHWIDE}}
\newcommand{\Ariadne}{\mbox{ARIADNE}}

\newcommand{\grcff}{\mbox{grc4f}}
\newcommand{\Vermaseren}{\mbox{VERMASEREN}}

\newcommand{\PhysLett}  {Phys.~Lett.}

\newcommand{\PhysRev}   {Phys.~Rev.}

\newcommand{\NIM} {Nucl.~Instr.\ and Meth.}
\newcommand{\ZPhys}  {Z.~Phys.}

\newcommand{\CPC} {Comp. Phys. Comm.}
\newcommand{\EPJ} {Eur.~Phys.~J.} 
\newcommand{\OPALColl}    {OPAL Collab.}
\documentclass[12pt]{article}
\usepackage{epsfig}
\usepackage{a4p}
\usepackage{amsbsy}
\usepackage{pennames} 
\usepackage{cite}
\begin{document}
\begin{titlepage}

\begin{center}{\large   EUROPEAN ORGANIZATION FOR NUCLEAR RESEARCH
}\end{center}\bigskip
\begin{flushright}
       CERN-PH-EP/2004-042   \\ 12th August 2004
\end{flushright}
\bigskip\bigskip\bigskip\bigskip\bigskip
\begin{center}
\huge\bf 
Determination of the \\ LEP Beam Energy 
using Radiative Fermion-pair Events 
\end{center}\bigskip\bigskip
\begin{center}{\LARGE The OPAL Collaboration
}\end{center}\bigskip\bigskip
\bigskip
\begin{abstract}
We present a determination of the \LEP\ beam energy using 
``radiative return'' fermion-pair events recorded at centre-of-mass  
energies from 183~GeV to 209~GeV\@. We find no evidence of a 
disagreement between the \Opal\ data and the \LEP\ 
Energy Working Group's standard calibration.
Including the energy-averaged $11$~MeV uncertainty in the standard 
determination,
the beam energy we obtain from the \Opal\ data is higher than
that obtained from the \LEP\ calibration by
\[ 
 0 \pm 34 (\mathrm{stat.}) \pm 27 (\mathrm{syst.})
\;\mathrm{MeV}.\]
\end{abstract}

\bigskip

\begin{center}

\vspace{2cm}

{\large (To be submitted to Physics Letters B)}

\end{center}
\end{titlepage}
\begin{center}{\Large        The OPAL Collaboration
}\end{center}\bigskip
\begin{center}{
G.\thinspace Abbiendi$^{  2}$,
C.\thinspace Ainsley$^{  5}$,
P.F.\thinspace {\AA}kesson$^{  3,  y}$,
G.\thinspace Alexander$^{ 22}$,
J.\thinspace Allison$^{ 16}$,
P.\thinspace Amaral$^{  9}$,
G.\thinspace Anagnostou$^{  1}$,
K.J.\thinspace Anderson$^{  9}$,
S.\thinspace Asai$^{ 23}$,
D.\thinspace Axen$^{ 27}$
% ...
}\end{center}\bigskip
\bigskip
$^{  1}$School of Physics and Astronomy, University of Birmingham,
Birmingham B15 2TT, UK
\newline
$^{  2}$Dipartimento di Fisica dell' Universit\`a di Bologna and INFN,
I-40126 Bologna, Italy
\newline
$^{  3}$Physikalisches Institut, Universit\"at Bonn,
D-53115 Bonn, Germany
\newline
% ...
$^{  5}$Cavendish Laboratory, Cambridge CB3 0HE, UK
\newline
% ...
$^{  9}$Enrico Fermi Institute and Department of Physics,
University of Chicago, Chicago IL 60637, USA
\newline
% ...
$^{ 16}$Department of Physics, Schuster Laboratory, The University,
Manchester M13 9PL, UK
\newline
% ...
$^{ 22}$Department of Physics and Astronomy, Tel Aviv University,
Tel Aviv 69978, Israel
\newline
$^{ 23}$International Centre for Elementary Particle Physics and
Department of Physics, University of Tokyo, Tokyo 113-0033, and
Kobe University, Kobe 657-8501, Japan
\newline
% ...
$^{ 27}$University of British Columbia, Department of Physics,
Vancouver BC V6T 1Z1, Canada
\newline
% ...
$^{  y}$ now at CERN
\newline
% ...

\newpage

\section{Introduction}

The measurement of the mass of the W boson, \MW, is one of the
principal goals of the \Leptwo\ program.  
The resolution on the measured W mass is greatly improved by 
employing kinematic fits, in which the constraints of energy and 
momentum conservation are imposed~\cite{yellowbook}. 
An accurate determination of the \LEP\ beam energy 
is therefore of paramount importance, since it sets the scale 
for the W mass measurement.  

The standard approach used to determine the average beam energy at 
\Leptwo~\cite{lepewg} 
involves precise measurements based on resonant depolarisation of the 
beams at energies in the range 41--61~GeV, combined with magnetic 
extrapolation to higher energies using \NMR\ probe measurements.
Corrections are applied to account for
variations of the beam energy with time, and for differences at the
four experimental interaction points around the ring. The \LEP\ Energy
Working Group calculates the beam energy for each experiment for
periods of 15 minutes, or more frequently if a change in 
operating conditions causes an abrupt shift in the beam energy. 
The systematic uncertainty in the beam energy is dominated by the
precision of approximately 10~MeV in the magnetic extrapolation and, 
uniquely in 2000, by the error of approximately 15~MeV associated with the 
strategy (so-called Bending Field Spreading) to boost the beam energy to 
the highest possible value. 

In this paper we assume the
modelling of variations in the \LEP\ beam energy to be correct and perform 
a check on the overall energy scale using radiative return events
of the type
\[ \Pep\Pem \rightarrow \PZ \Pgg; \;\;\;
\PZ \rightarrow \mathrm{f\overline{f}},\]
where the fermion f is a quark, electron, muon or $\tau$-lepton.
Since the Z mass is very precisely known from \Lepone~\cite{pdg}, the
kinematic properties of these events can be used to estimate the beam energy.
For hadronic events, information is taken from the jet energies
and directions, while for leptonic events only the angular
information is used.

The results of these measurements of the beam energy, using the 
information from observed events, can be interpreted in several ways. 
Any discrepancy could indicate a problem with the 
\LEP\ energy calibration. Alternatively, since the techniques employed are 
closely related to those used in the W mass measurement, they could be 
regarded as a check of detector systematic errors, or of hadronisation
uncertainties in the case of hadronic events.  The results could also be 
regarded as a check on the Monte Carlo 
modelling of initial-state radiation (ISR) 
in the radiative return process. 
 
This paper is organised as follows: a summary of the data and Monte Carlo
samples used is given in Section~\ref{sect-datamc}, the analysis method is
explained in  Section~\ref{sect-anal} and the estimation of systematic errors
is described in Section~\ref{sect-syst}.  Finally we summarise and discuss 
the results in Section~\ref{sect-summ}.

\section{Data and Monte Carlo Samples}
\label{sect-datamc}

The \Opal\ detector\footnote{\Opal\ uses a right-handed coordinate system in
which the $z$ axis is along the electron beam direction and the $x$
axis is horizontal. The polar angle $\theta$ is measured with respect
to the $z$ axis and the azimuthal angle $\phi$ with respect to the
$x$ axis.}, trigger and data acquisition system are fully described 
elsewhere~\cite{bib:OPAL-detector,bib:OPAL-SI,bib:OPAL-SW,bib:OPAL-TR,
bib:OPAL-DAQ}.  The data used for the present analysis were collected between
1997 and 2000, at centre-of-mass energies in the range from 183~GeV to
209~GeV. The approximate amount recorded at each nominal energy is given in
Table~\ref{tab-lumi}. 

\begin{table}[htb]
\begin{center}
\begin{tabular}{ccc}
\hline
Year(s) & $\sqrt{s}$ /GeV & $\int\!{\cal L}\mathrm{d}t$ /pb$^{-1}$  \\ 
\hline
1997      &     183          &  58  \\
1998      &     189          & 186  \\
\hline
1999      &     192          &  30  \\
1999      &     196          &  78  \\
1999+2000 &     200          &  79  \\
1999+2000 &     202          &  38  \\
2000      &     205          &  82  \\ 
2000      &     207          & 137  \\
\hline
1999      &   192--202       & 223  \\
2000      &   200--207       & 221  \\  
\hline
\end{tabular}
\end{center}
\caption{\sl Nominal centre-of-mass energies and approximate
integrated luminosities for data collected between 1997 and 2000.}
\label{tab-lumi}
\end{table}

Samples of Monte Carlo simulated events are used to interpret the data.  
Separate Monte Carlo samples were generated at each of the 
nominal centre-of-mass energy values considered and also at several 
intermediate points. The programs employed for this purpose are outlined below.
First we give those used to generate signal events, then those for generation
of the various backgrounds. All Monte Carlo samples were passed through the 
\Opal\ detector simulation program~\cite{bib:GOPAL}, and processed 
in the same way as real data.

For the hadronic final states, the $\KK$~\cite{bib:KK2f} program (v.4.13) 
is used to generate the $\mathrm{q\overline{q}}(n\Pgg)$ process (where $n$ is
an integer), including the signal $\mathrm{q\overline{q}}\Pgg$ events, and 
likewise the $\mu^+\mu^-(n\Pgg)$ and $\tau^+\tau^-(n\Pgg)$ processes. In this 
scheme, ISR is modelled with Coherent Exclusive Exponentiation 
(\CEEX)~\cite{bib:ceex} to ${\cal{O}}(\alpha^2)$ precision. 
For the $\epem(n\Pgg)$ final-state process,
\Bhwide~\cite{bib:bhwide} (v.1.00) is employed, in which ISR is modelled with 
YFS~\cite{bib:yfs} exponentiation to ${\cal{O}}(\alpha)$ precision.
For the hadronic final states,
fragmentation of the primary quarks is performed according to the
\Pythia\ (v.6.150)~\cite{bib:PYTHIA} prescription, 
with \Herwig\ (v.6.2)~\cite{bib:HERWIG}
and \Ariadne\ (v.4.11)~\cite{bib:ARIADNE} employed 
for systematic studies.
In order to simulate properly the interplay between photon and gluon radiation
in the final-state parton shower, final-state radiation (FSR) of photons is 
turned {\em off}\/ in the generation of the primary quark pairs in $\KK$ 
and turned {\em on}\/ in the hadronisation programs. As a consequence, the
Monte Carlo for hadronic events does not include the interference between
initial- and final-state photon radiation (I/FSR interference) which is 
naturally present in the
data. The absence of this is taken as a systematic uncertainty, as described in
Section~\ref{sect-syst_h}. For leptonic final states this problem does not
arise and the Monte Carlo includes I/FSR interference.
  
Four-fermion backgrounds are simulated using \grcff~\cite{bib:GRC4F}
or \Koralw~\cite{bib:KORALW} with matrix elements from \grcff, and
two-photon backgrounds using \Phojet~\cite{bib:phojet},
\Pythia, \Herwig, \Twogen~\cite{bib:twogen}
and \Vermaseren~\cite{bib:verm}. For systematic studies of tau-pair backgrounds
in the hadronic channel, the \Koralz~\cite{bib:KORALZ} generator is also
employed.

\section{Analysis Method}
\label{sect-anal}
\subsection{Hadronic Channel}
\label{sect-anal_h}
\subsubsection{Event Selection}

The analysis in the hadronic final state closely follows the procedures 
used in the measurement of hadronic 
cross-sections~\cite{bib:cga_thesis,bib:OPAL-SM209,bib:OPAL-SM189,
bib:OPAL-SM183,bib:OPAL-SM172}---hadronic events are selected according to the 
same criteria and the effective centre-of-mass energy of the hadronic system
after ISR, $\sqrt{s'}$, is computed by an identical algorithm.
In summary, the algorithm to determine $\sqrt{s'}$ starts by identifying 
isolated photons in the electromagnetic calorimeter with 
energies greater than 10~GeV\@, based on their expected narrow lateral shower
shapes and their lack of penetration into the hadronic calorimeter.
The remaining tracks and clusters (in both
electromagnetic and hadronic calorimeters) are formed 
into jets using the Durham algorithm~\cite{Durham} with a jet resolution
parameter $\ycut=0.02$. If more than four jets are found, a four-jet 
configuration is enforced nevertheless.
A standard algorithm~\cite{bib:MT} is applied
to reduce double counting of energy before calculating
the jet energies, masses and directions. 
As was done for the W mass analysis~\cite{bib:OPAL-MW189},
small corrections to the jet parameters and their errors are
applied to improve the consistency between data and Monte Carlo, 
based on studies of Z calibration data and of full-energy events in the high
energy data. A kinematic fit
is performed to improve the estimates of the jet four-momenta by
imposing the constraints of energy and momentum conservation.
The r\^{o}le of the beam energy in this fit is elaborated on below.
If this fit is unsuccessful, 
an additional unseen photon is assumed moving parallel to 
the beam direction ($z$), and the kinematic fit is repeated.  If this fails, 
a fit involving two unmeasured photons in the $\pm z$ directions is attempted.
The value of $\sqrt{s'}$ is obtained as the invariant mass of the 
jets resulting from the first successful fit. Events classified by the
algorithm as having exactly one photon, either measured in the calorimeter 
or parallel to $z$, are retained for analysis; events classified as 
having multiple photons are discarded, suffering from poorer resolution on 
$\sqrt{s'}$ or higher background. The typical resolution on
$\sqrt{s'}$ is around 2~GeV, though with tails associated with 
unresolved multiple soft photon radiation.  

\subsubsection{Fitting Method and Results}

\begin{figure}
  \epsfxsize=\textwidth
  \epsffile{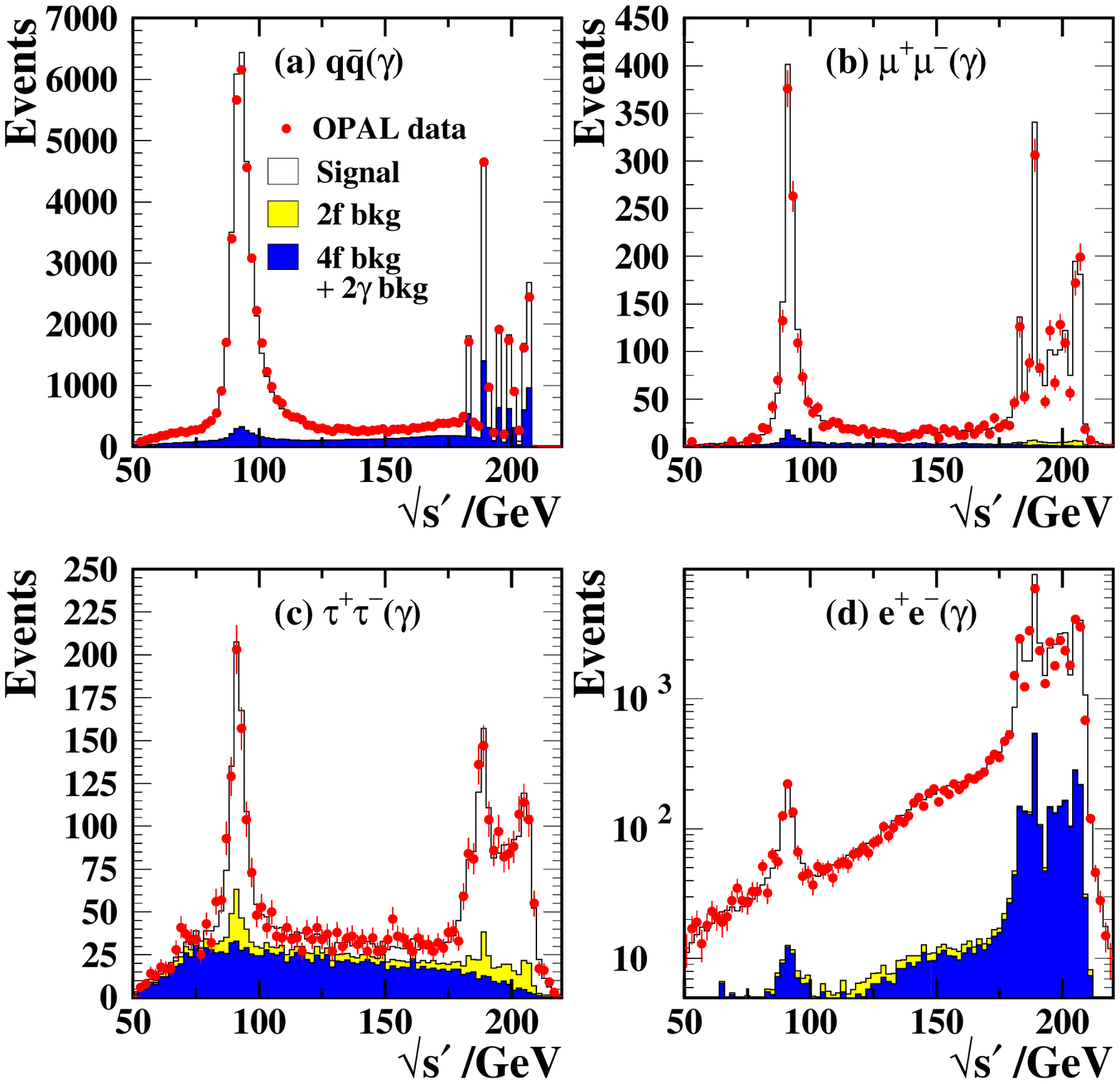}
  \caption{\sl Distributions of $\sqrt{s'}$ for (a) hadronic 
and (b)--(d)
leptonic events before applying cuts on photon radiation. Data 
with centre-of-mass energies between 183~GeV and 209~GeV have been combined.
Full-energy events from data-taking at 
different centre-of-mass energies are responsible
for the multiple peaks observed at high $\sqrt{s'}$.
The corresponding Monte Carlo expectation is also shown, normalised to the 
integrated luminosity of the data. 
The Monte Carlo samples are not generated at exactly the same energies
as the data, which together with binning effects explains the visible
differences in structure for full-energy events.
(The poorer resolution for tau-pair events washes out this effect.)
The radiative return peak is dwarfed by the contribution from $t$-channel 
full-energy events for electrons.}
\label{fig:eb_spr}
\end{figure}

The reconstructed $\sqrt{s'}$ distributions of the data and Monte Carlo 
are compared for hadronic events in Fig.~\ref{fig:eb_spr}(a); 
the Z mass peak is clearly seen.
The background to the $\PZ\Pgg$ final state is around 4\%, and is dominated
by the $\mathrm{q\overline{q}e^+e^-}$ four-fermion process in which the 
$\mathrm{q\overline{q}}$ arise from the decay of a Z boson, so most of
these events can also effectively be regarded as signal.   
The calculation of $\sqrt{s'}$ relies on 
the constraint in the kinematic fit that the energies of the jets and 
photons add up to the centre-of-mass energy. In Monte Carlo events,
the correct centre-of-mass energy is of course known {\em a priori}.  
In the case of data, we use the beam energies determined from the magnetic 
extrapolation method by the \LEP\ Energy Working Group\@.
Any systematic inaccuracy in this estimate of the beam energy would be 
manifested as a shift in the reconstructed Z peak in data.  
The basis of the analysis method is therefore to reconstruct $\sqrt{s'}$ in 
the data as a function of an assumed difference, $\Delta\Ebeam$,
between the real beam energy and that estimated from magnetic extrapolation, 
and to find the value of $\Delta\Ebeam$ which optimises the 
agreement between the peaks in data and Monte Carlo\@. The sign of 
$\Delta\Ebeam$
is such that a positive (negative) value implies that the value determined 
from \Opal\ data is higher (lower) than that determined by \LEP\@.  

To compare data and Monte Carlo, we fit an empirical analytic function to the 
Z mass peak for 26 bins in the region $87<\sqrt{s'}/\mathrm{GeV}<100$ and 
characterise the distributions by the fitted peak position, $M^*$. 
The function chosen has the form
\begin{eqnarray}
S(\sqrt{s'}) & = & A\left[
   c\left(\frac{2\sqrt{s}}{s-s'}\right) 
   \frac{s'\Gamma_{\pm}^{2}}{(s'-{M^*}^2)^{2}+s'\Gamma_{\pm}^{2}}
   +a(1+b\sqrt{s'})\right],
\label{eq_rbw}
\end{eqnarray}
where
\begin{eqnarray}
\Gamma_{\pm} & = & \left\{ \begin{array}{ll}
                   \Gamma_-, 
                 & \mathrm{for}\ \sqrt{s'} < M^*; \\
                   \Gamma_+, 
                 & \mathrm{for}\ \sqrt{s'} > M^*. \\
                  \end{array}
          \right. \nonumber
\end{eqnarray}
It consists of two parts that, together, are found to fit the peak well. 
The first part describes the contribution of
processes which are resonant at the Z peak, including the signal $\qq\gamma$
production. It comprises a pair of matched 
relativistic Breit-Wigner functions with different widths, 
$\Gamma_{-}$ and $\Gamma_+$, below and 
above the peak respectively, and a normalisation factor, $c$. The factors
of $\Gamma_{\pm}^2$ in the numerator ensure continuity of the function 
at $\sqrt{s'}=M^*$. The factor $\frac{2\sqrt{s}}{s-s'}$ is intended to 
represent the effect of a bremsstrahlung spectrum proportional to the 
reciprocal of the energy of an ISR photon, though 
it actually has a rather small effect. The second part describes the 
non-resonant background contribution. It is a function linear in
$\sqrt{s'}$, with a parameter, $b$, determining 
the shape and a parameter, $a$, 
providing normalisation. First of all the background parameters are extracted 
from fits to Monte Carlo simulations of two-photon and four-fermion (excluding
$\mathrm{q\overline{q}}\Pep\Pem$) events, which are non-resonant under the
peak, at several centre-of-mass energies; their energy-dependences are 
taken from linear fits. The parameters $\Gamma_{\pm}$ and $c$ are then 
extracted from fits to Monte Carlo, including both the resonant and the 
non-resonant contributions, at the same centre-of-mass energies, with the
background parameters constrained to those previously determined; their
energy-dependences are also taken from linear fits. Finally, in fitting the 
Monte Carlo and data with all parameters constrained to their energy-determined
values, only the overall normalisation, $A$, and the peak position, $M^*$, are 
allowed to vary. 

\begin{figure}
  \epsfxsize=\textwidth
  \epsffile{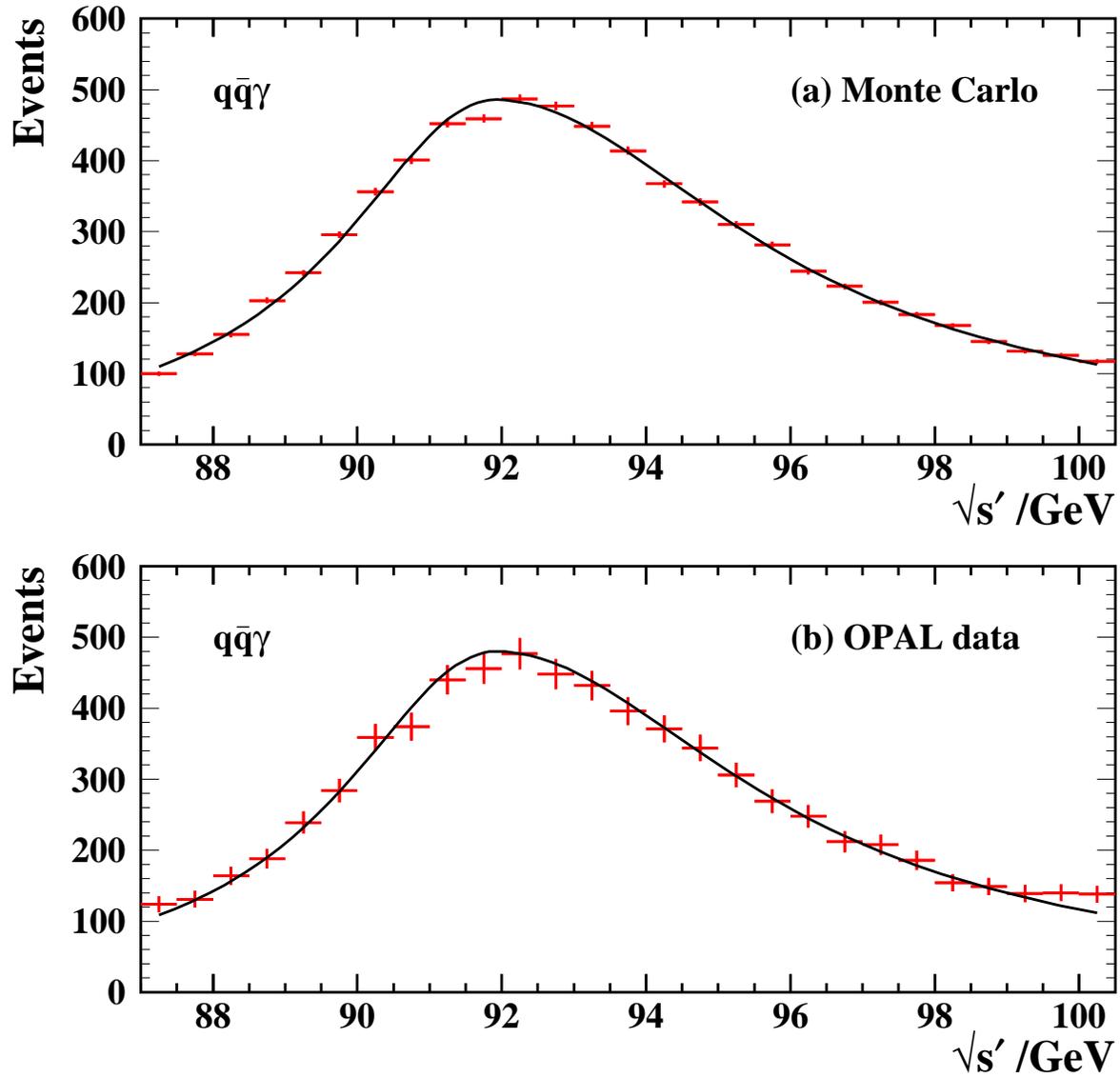}
  \caption{\sl Fits of Eq.~(\ref{eq_rbw}) to (a) Monte Carlo 
generated at 189~GeV and (b) \Opal\ data collected in 1998, at the 
same nominal energy, for hadronic events. The Monte Carlo expectation is 
normalised to the integrated luminosity of the data.}
\label{fig:spr_hadfits}
\end{figure}

\begin{figure}
  \epsfxsize=\textwidth
  \epsffile{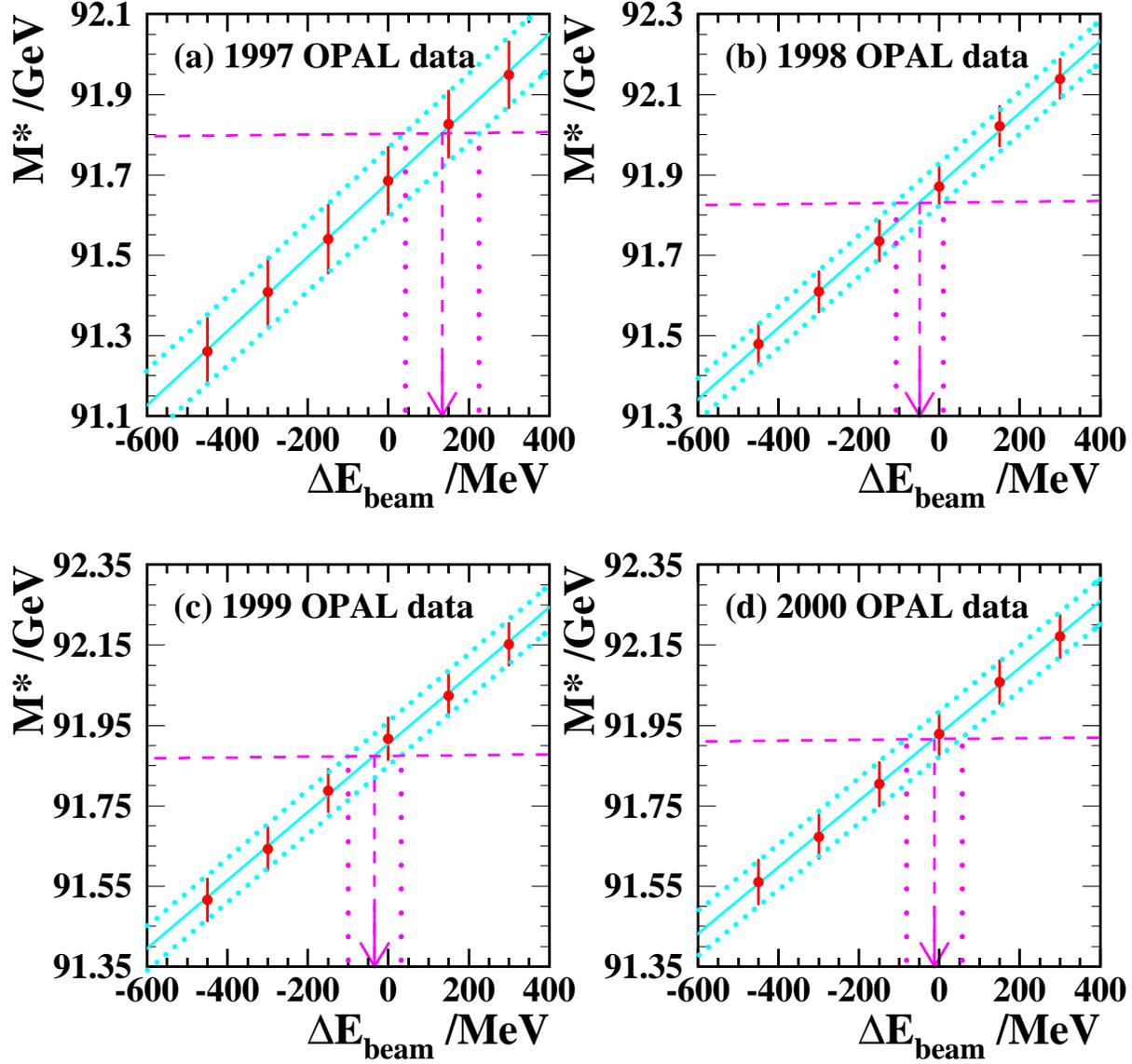}
  \caption{\sl Extraction of $\Delta\Ebeam$ from 
hadronic events in \Opal\ data collected during 
(a) 1997, (b) 1998, (c) 1999 and (d) 2000\@. 
Each plot shows the value of the peak position, $M^*$, obtained 
from data as a function of the assumed correction to the \LEP\ beam energy, 
$\Delta\Ebeam$\@. The solid line is a fit to the points, while the diagonal 
dotted lines define the statistical error band. The near-horizontal dashed 
line indicates the Monte Carlo expectation for
$M^*$ as a function of $\Delta\Ebeam$\@. The intersection of this with the 
diagonal band allows the true value of $\Delta\Ebeam$ and its statistical 
error to be inferred from the data.}
\label{fig:eb_extract}
\end{figure}

From the data recorded in the years 1997, 1998, 1999
and 2000, the numbers of selected events in the fit region are 2386, 
7238, 7198 and 6300 respectively.
Typical fits used to determine $M^*$ in Monte Carlo and data 
are shown in Fig.~\ref{fig:spr_hadfits}.  
The method for estimating $\Delta\Ebeam$
is illustrated by  Fig.~\ref{fig:eb_extract}, which shows 
the value of $M^*$ obtained from the data as a function of 
the assumed value of $\Delta\Ebeam$.  The data points define a 
band of constant width, since the statistical errors 
are almost fully correlated from point to point.
The fitted value of $M^*$ in the Monte Carlo is
evaluated for a range of generated beam energies; a weak dependence
of about 10 MeV in $M^*$ for a 1 GeV change in beam energy is observed. To
account for this, as the beam energy (and hence $\Delta\Ebeam$) is varied in 
the data, the corresponding value of
$M^*$ in data is compared with the expected value of $M^*$ in Monte Carlo
for a {\em known}\/ beam energy in Monte Carlo corresponding to this 
{\em assumed}\/ average beam energy in data. This is represented by the 
near-horizontal line.
The point where this crosses the data therefore gives the 
estimated value of $\Delta\Ebeam$ in the data.
      
The values of $\Delta\Ebeam$ with their statistical errors
are given in Table~\ref{tab:results-all}. The systematic errors are 
estimated as described in Section~\ref{sect-syst_h} below. Measurements on
subsets of the data collected at the nominal energy points detailed in 
Table~\ref{tab-lumi} are performed in an equivalent manner.

\subsection{Leptonic Channels}
\label{sect-anal_l}
\subsubsection{Event Selection}
Although the leptonic channels offer lower statistics than
the hadronic final state, the systematic uncertainties
associated with the measurement are different.
Of the three lepton species, the muon sample
gives the most precise result, benefitting from a 
very low background and an
excellent angular resolution for the two muons. 
The tau channel suffers from lower selection efficiency, 
a worse resolution of the tau-lepton 
direction and a larger background. 
The situation for the electron channel is 
complicated by the 
$t$-channel exchange contribution.
Nonetheless, this channel turns out to be more precise than
the tau channel.

In general, the lepton-pair
event selection looks for the two charged leptons, and
possibly a photon in the detector. Photons are identified 
as clusters in the electromagnetic calorimeter with a 
narrow shower shape consistent with being a photon, no
associated track and with energy 
greater than 5~GeV. Only the
highest energy photon candidate is considered. In
all cases the event is treated as having exactly three final-state particles,
two leptons and a photon. 
If no photon is observed, then the third particle is
taken to be a photon along the beam axis, recoiling against the
two-lepton system. Events with an observed
photon between 5~GeV and 30~GeV are rejected because they
would fall far away from the
radiative return peak if there were only one final-state photon
in this energy range.
Events with an observed photon with
energy greater than 30~GeV
are assumed to have no radiation along the beam
direction. The angles of all tracks, and of calorimeter clusters in 
the electron-pair events, are calculated taking into account 
the offset of the beam spot position from the nominal
detector origin.

The planarity of the event is defined as the sum of the
three angles between the directions of the two leptons, and 
the direction of the photon, either the observed photon
in the detector or along the beam axis.
The event planarity must be
greater than $350^\circ$. True three-body events and events with only
collinear initial-state radiation along the $z$-axis are planar,
unlike events from four-fermion processes, for example.

The selection of muon events used here is the same as in
Refs.~\cite{bib:OPAL-SM189,bib:OPAL-SM183,bib:OPAL-SM172},
with the addition of the planarity cut defined above.
A total of 3604
muon-pair events is selected in the combined 1997 to 2000 data,
with 9\% background according to the Monte Carlo. There
are 1166 events in the radiative return peak region, 
defined by $82<\sqrt{s'}/\mathrm{GeV}<102$,
with 6\% background, dominated by four-fermion and
two-photon processes.

Although the tau sample includes a larger background than the
muon channel, background from other processes
including Z decay to a fermion pair can be included in the signal, while
background from two-photon and other four-fermion processes is flat
under the radiative return peak. A dedicated tau selection is therefore
used here, which is somewhat more efficient than that used in the
\Opal\ two-fermion cross-section analyses, at the expense of including more
background.

The tau selection excludes events which are 
identified as $\Pep\Pem(\gamma)$ or $\mu^+\mu^-(\gamma)$ candidates.
Low multiplicity events are required,
with at least 2 and not more than 6 tracks.
The number of tracks 
plus the number of energy clusters
in the electromagnetic calorimeter must be less than 16.
The events are required to be consistent with originating from
the beam collision in space and time
to reject cosmic rays.
A cone jet finding algorithm is applied, searching for 
cones of half angle $45^\circ$, each containing at least 5\% of the
centre-of-mass energy. At least two cones must
be identified in the event. If only two cones are found, 
these are assumed to be $\tau$-leptons, with an 
unobserved photon in the beam pipe.
If three or more cones are found, then isolated photon
candidates with energy greater than 5~GeV
are also identified. The cone containing
the highest energy photon candidate is taken to be
the photon when reconstructing the event. 
Of the remaining cones, the two with highest
energy are taken to be the $\tau$-leptons. If there are 
three or more cones but no identified photon, then
the two highest energy cones are taken to be the $\tau$-leptons,
and the third most energetic cone is treated as an 
observed photon. Events 
where the ``photon'' cone has energy less than 30~GeV are
then rejected.

The two tau cones must satisfy
\[ | \cos \theta_{\mathrm{cone}} | < 0.9 \]
to reduce the contamination from $t$-channel Bhabha events.
The scalar sum of the energy in tracks and clusters
(with no correction for double counting) divided
by the centre-of-mass energy must be in the range
\[   0.3 < \Sigma E / \sqrt{s} < 1.1 \;\;\;.\]
Low energy events are predominantly from two photon events,
and high energy events are overwhelmingly dominated by Bhabhas.

In the combined 1997 to 2000 data, 4173 events are selected in the 
data, which according to the Monte Carlo comprise 52\% tau pairs, with
an additional 12\% of ``resonant'' background ({\em i.e.}\/ electron-,
muon- and quark-pair events). Under the radiative return peak,
there are 973 events, and 
the purity increases to 59\% tau pairs, with 
a further 12\% resonant events.
The average efficiency for selecting tau-pair events
over all the centre-of-mass energies
studied is 46\%.

The selection of $\Pep\Pem(\gamma)$ events used here is almost entirely based
on calorimeter information, motivated by avoiding systematic
uncertainties in the modelling of forward, high-energy electron tracks.
Low multiplicity events are required, with fewer than 18 tracks
plus clusters, 
and events selected as muon pairs are excluded.
Photon or electron-like clusters are
identified in the electromagnetic calorimeter
by applying the photon selection,
but allowing tracks to be associated
with the clusters. These clusters are sorted according to
their energy.
The two or three most energetic clusters must satisfy 
$ E_1 + E_2 + E_3 > 0.7 \sqrt{s} /2 $ and 
$E_2 > 0.2  \sqrt{s}/2 $.  
The two highest energy
clusters must be in the angular region
$| \cos \theta_{\mathrm{clus}} | <0.9$. 
The same scalar sum of energy in tracks and clusters as used above in
the tau-channel selection must pass $ \Sigma E / \sqrt{s} > 1.0 $.

In the 1997 to 2000 data, a total of 47,775 events is selected. This
number is overwhelmingly dominated by nearly back-to-back $t$-channel
exchange events, with only 825 events falling in the radiative
return peak region. Under the peak, 1\% of the events 
are from resonant backgrounds 
and 9\% from other backgrounds, dominated by four-fermion
and two-photon processes. (The $s$- and $t$-channel processes
are not separated---all $\Pep\Pem(\gamma)$ 
final states count as ``signal''.)

\subsubsection{Fitting Method and Results}

Each leptonic event is assumed to contain exactly three
final-state particles: two leptons plus
one and only one photon. 
The ratio of $s'/s$ 
is determined from the directions of these particles.
The photon is assumed to travel along the beam axis,
recoiling against the leptons, unless a photon candidate 
with energy greater than 30~GeV 
is observed in the detector, in which case the
direction of the electromagnetic cluster is taken to
be the photon direction.
For muon-pair events, the directions of the muon tracks are used, and
for tau pairs the directions of the cones, defined as the vector sum of
the tracks and clusters in the cone, without any correction for
double counting. For electrons, the directions of the electromagnetic
calorimeter clusters are taken.
The only energy information used is the loose 30~GeV
minimum energy requirement on an observed photon. 

The ratio $s'/s$ is given by
\begin{eqnarray}
\label{eq:spos}
\frac {s'}{s} &  = &  
\frac
{ \sin \alpha_1 + \sin \alpha_2 - | \sin (\alpha_1 + \alpha_2) | }
{ \sin \alpha_1 + \sin \alpha_2 + | \sin (\alpha_1 + \alpha_2) | },
\end{eqnarray}
where $\alpha_1$ and $\alpha_2$ are taken to be 
the polar angles $\theta_{1,2}$ of the two leptons in the detector 
if the photon 
is undetected, or the angles between the two leptons and the photon direction
if the photon is detected. 
The distributions of $\sqrt{s'}$ are shown in 
Figs.~\ref{fig:eb_spr}(b), (c) \& (d) for the muon, tau and electron samples
respectively.
For genuine radiative return events, the value of
$\sqrt{s'}$ is approximately equal to the mass of the Z boson.

As for the hadronic samples,
the values of $\sqrt{s}$ for the data are those provided
by the \LEP\ Energy Working Group, while for the Monte Carlo
sample, the true value is known exactly from the event generation.
Fits are made to the $\sqrt{s'}$ distribution 
for 20 bins in the region $82<\sqrt{s'}/\mathrm{GeV}<102$ for the 
muon and tau-pair samples, and 40 bins spanning 72~GeV to 112~GeV for
the electron-pair events,
to allow the $t$-channel contribution to be
constrained by the data. For muon and tau events, the same method is used
as for hadrons, but with the parameter $b$ in Eq.~(\ref{eq_rbw})
set to zero so that the non-resonant background is assumed
to be constant.
For the electron sample, a first fit for the parameter $a$ 
is made as before to the non-resonant background Monte Carlo samples alone.
Then, when fitting the signal plus background,
an additional linear term of the 
form $ f(1+g\sqrt{s'})$
is included to account for the $t$-channel 
contribution to the signal. The parameters $f$ and $g$ are similarly 
fixed from the Monte Carlo simulation. 
Data and Monte Carlo samples
from all centre-of-mass energies are fitted together for
the central result, allowing for no energy dependence of
the parameters describing the background, the widths or
the value of $M^*$.
Separate fits are also made for data from each year
of running, 
using Monte Carlo samples from the same range of centre-of-mass
energies.

\begin{figure}
  \epsfxsize=\textwidth
  \epsffile{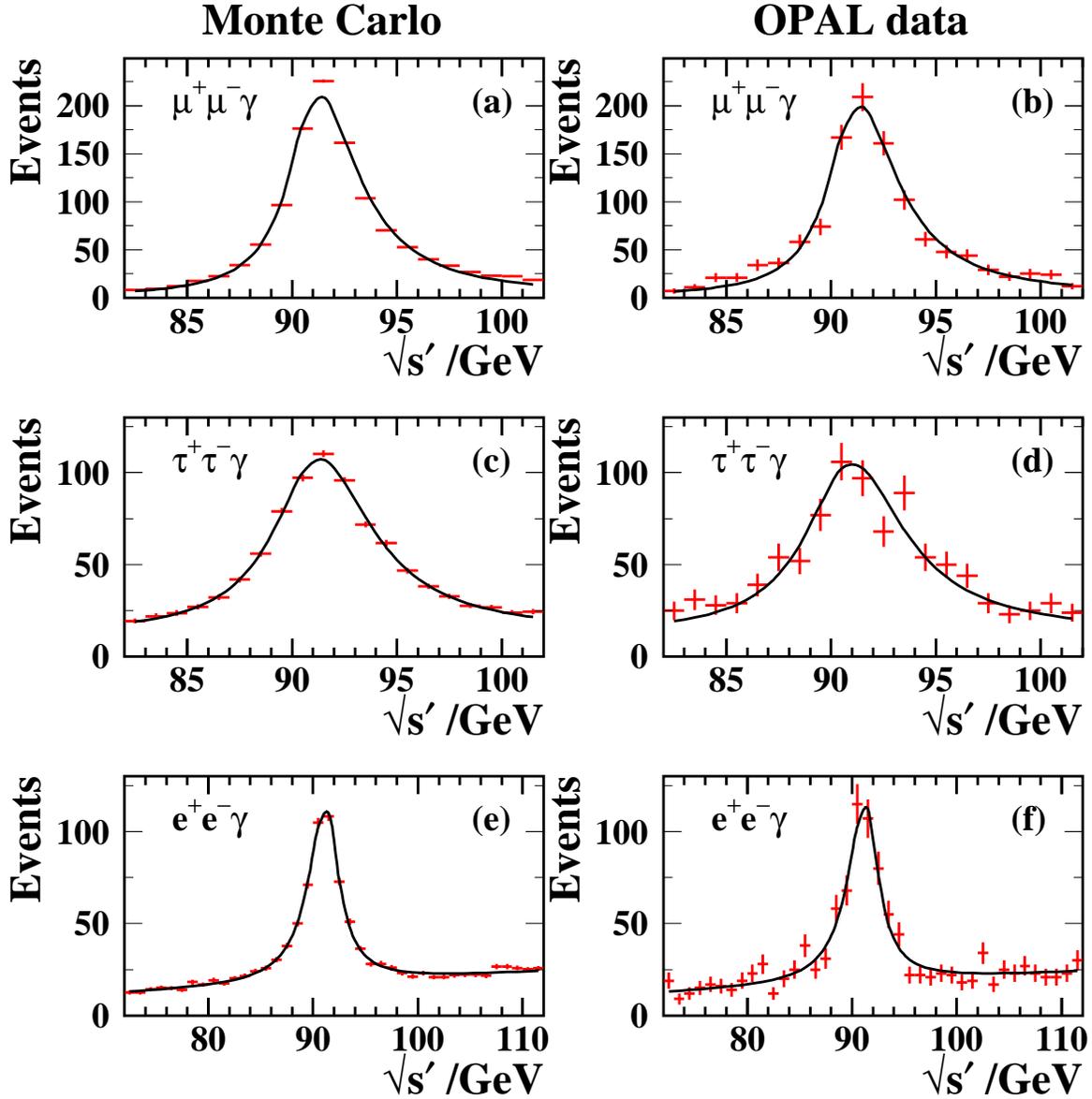}
  \caption{\sl Fits of Eq.~(\ref{eq_rbw}) to Monte Carlo (left-hand plots), 
combining
samples generated at energies in the range 183--209~GeV, 
and \Opal\ data (right-hand plots) collected in the years
1997--2000 at the same nominal energies, for muon-, tau- and electron-pair 
events respectively. 
The Monte Carlo expectation is normalised to the integrated 
luminosity of the data in each case.}
\label{fig:leptonfits}
\end{figure}

The variation of the position of the peak $M^*$ in the Monte
Carlo is evaluated
as a function of a shift in the assumed beam energy $\Delta\Ebeam$. This
is used to convert the difference between the peaks in 
data and Monte Carlo into the difference between the beam
energies observed in \Opal\ and provided by the \LEP\ Energy Working Group.
The results are given in 
Table~\ref{tab:results-all},
and the data quality is illustrated 
in Fig.~\ref{fig:leptonfits}.

Cross-checks are made using different fitting methods.
Fits are made to the distribution of reconstructed
centre-of-mass energy, found using Eq.~\ref{eq:spos},
assuming that $s' \equiv \MZ$ in every event.
The binned data and Monte Carlo
distributions are also compared directly as a function
of the beam energy offset, instead of using an
empirical functional form.
In each case consistent results are found. 

\newcommand {\wpm}{$\!\!\!\!\!\!\pm\!\!\!\!\!\!$}
\newcommand {\wbl}{$\!\!\!\!\!\!   \!\!\!\!\!\!$}

\begin{table}[htb]
\begin{center}
\begin{tabular}{crccclrccclrccclrcccl|rcccl}
\hline
                     Year & \multicolumn{25}{c}{$\Delta\Ebeam$ /MeV}  \\
\cline{2-26} 
                          & \multicolumn{5}{c}{$\qq\gamma$}
                          & \multicolumn{5}{c}{$\mumu\gamma$}
                          & \multicolumn{5}{c}{$\tautau\gamma$}
                          & \multicolumn{5}{c|}{$\epem\gamma$}
                          & \multicolumn{5}{c}{All channels}          \\
\hline
              1997   &  $\!\!+$134 & \wpm &   92 & \wpm & 33$\!\!$ & 
                        $\!\!+$577 & \wpm &  251 & \wpm & 29$\!\!$ & 
                      $\!\!+$1\,157 & \wpm &  548 & \wpm & 89$\!\!$ & 
                      $\!\!-$1\,590 & \wpm &  589 & \wpm & 86$\!\!$ & 
                        $\!\!+$176 & \wpm &   84 & \wpm & 28$\!\!$ \\ 
              1998   &   $\!\!-$49 & \wpm &   59 & \wpm & 52$\!\!$ & 
                         $\!\!+$71 & \wpm &  133 & \wpm & 30$\!\!$ & 
                        $\!\!+$266 & \wpm &  282 & \wpm & 93$\!\!$ & 
                        $\!\!+$172 & \wpm &  217 & \wpm & 75$\!\!$ & 
                          $\!\!+$8 & \wpm &   53 & \wpm & 37$\!\!$ \\ 
\hline
             192 GeV &  $\!\!-$103 & \wpm &  123 & \wpm & 30$\!\!$ & 
                               & \wbl &   -- & \wbl &    &   
                               & \wbl &   -- & \wbl &    &
                               & \wbl &   -- & \wbl &    &
                               & \wbl &   -- & \wbl &    \\
             196 GeV &   $\!\!-$37 & \wpm &  117 & \wpm & 36$\!\!$ &
                               & \wbl &   -- & \wbl &    &   
                               & \wbl &   -- & \wbl &    &
                               & \wbl &   -- & \wbl &    &
                               & \wbl &   -- & \wbl &    \\
             200 GeV &   $\!\!+$35 & \wpm &  113 & \wpm & 37$\!\!$ &
                               & \wbl &   -- & \wbl &    &   
                               & \wbl &   -- & \wbl &    &
                               & \wbl &   -- & \wbl &    &
                               & \wbl &   -- & \wbl &    \\
             202 GeV &   $\!\!-$98 & \wpm &  183 & \wpm & 39$\!\!$ & 
                               & \wbl &   -- & \wbl &    &   
                               & \wbl &   -- & \wbl &    &
                               & \wbl &   -- & \wbl &    &
                               & \wbl &   -- & \wbl &    \\
             205 GeV &    $\!\!+$4 & \wpm &  106 & \wpm & 68$\!\!$ &  
                               & \wbl &   -- & \wbl &    &   
                               & \wbl &   -- & \wbl &    &
                               & \wbl &   -- & \wbl &    &
                               & \wbl &   -- & \wbl &    \\
             207 GeV &   $\!\!-$24 & \wpm &   93 & \wpm & 73$\!\!$ &
                               & \wbl &   -- & \wbl &    &   
                               & \wbl &   -- & \wbl &    &
                               & \wbl &   -- & \wbl &    &
                               & \wbl &   -- & \wbl &    \\
\hline
                1999 &   $\!\!-$34 & \wpm &   66 & \wpm & 36$\!\!$ & 
                         $\!\!-$71 & \wpm &  131 & \wpm & 28$\!\!$ & 
                        $\!\!+$529 & \wpm &  291 & \wpm & 88$\!\!$ & 
                        $\!\!-$271 & \wpm &  270 & \wpm & 70$\!\!$ & 
                         $\!\!-$30 & \wpm &   57 & \wpm & 27$\!\!$ \\ 
                2000 &   $\!\!-$12 & \wpm &   69 & \wpm & 72$\!\!$ & 
                        $\!\!-$293 & \wpm &  165 & \wpm & 33$\!\!$ & 
                        $\!\!+$399 & \wpm &  448 & \wpm &108$\!\!$ & 
                        $\!\!-$393 & \wpm &  303 & \wpm & 73$\!\!$ & 
                         $\!\!-$89 & \wpm &   65 & \wpm & 51$\!\!$ \\ 
\hline
           All years &    $\!\!+$1 & \wpm &   38 & \wpm & 40$\!\!$ & 
                         $\!\!-$32 & \wpm &   75 & \wpm & 25$\!\!$ & 
                        $\!\!+$313 & \wpm &  175 & \wpm & 76$\!\!$ & 
                         $\!\!-$88 & \wpm &  146 & \wpm & 46$\!\!$ & 
                          $\!\! $0 & \wpm &   34 & \wpm & 27$\!\!$ \\ 
\hline
\end{tabular}
\end{center}
\caption{\sl 
Summary of the values of $\Delta\Ebeam$ derived from hadronic and leptonic
events for each year and for all years combined. (For the statistically more
precise hadronic channel, the results are also presented at the individual 
nominal energy points for data collected in the years 1999 and 2000\@.) 
In each case, the first error is statistical and the second systematic.}
\label{tab:results-all}

\end{table}

\section{Systematic Errors}
\label{sect-syst}
\subsection{Hadronic Channel}
\label{sect-syst_h}

The evaluation of systematic errors closely follows the approach used in the 
measurement of hadronic cross-sections~\cite{bib:cga_thesis,bib:OPAL-SM209,
bib:OPAL-SM189,bib:OPAL-SM183,bib:OPAL-SM172}. The following effects are 
taken into account, and the uncertainties are summarised in 
Table~\ref{had_syst}.
\begin{itemize}

\item {\bf Detector modelling:} 
The inputs to the kinematic fits which 
determine $\sqrt{s'}$ are the measured energies, masses and angles of jets and 
photons and their resolutions.  For the measurement of the 
W mass~\cite{bib:OPAL-MW189}, studies of calibration data taken at the 
Z peak are used to apply small corrections to these energies and angles in 
the Monte Carlo simulation in order to achieve agreement with the data.
The same corrections, determined separately for each year of data-taking 
where appropriate, are applied in the present analysis. The 
errors in the correction factors are then taken to define 
systematic errors in the value of $\Delta\Ebeam$. Of particular concern are 
potential systematic shifts in the reconstruction of the polar angular scale, 
$\theta$, of jets (equivalent to an uncertainty in the effective 
length/radius ratio of the detector), as these could bias the reconstruction 
of $\sqrt{s'}$. These are assessed by comparing the jet angles in Z events 
reconstructed using tracking and calorimetry separately. 

In addition, the effects of deviations from linearity in the jet
energy scale of around $\pm1\%$, going from $\sim20$~GeV to
$\sim100$~GeV, are corrected for and the error in the correction is 
taken as a further source of systematic uncertainty. This non-linearity is 
assessed from studies of three-jet events in Z decays and of full-energy 
hadronic events in the high energy data. The linearity of the photon energy
scale is likewise studied using $\epem\gamma$ and $\mumu\gamma$ events in 
both the Z calibration data and the high energy data.
Though no significant deviations from linearity are seen in this case,
the error in the determination of the linearity is similarly used to define a 
systematic uncertainty. 

A further consideration is the uncertainty in the measured masses of jets.
Studies of Z calibration data suggest that the relationship between the jet 
mass scales in Monte Carlo and data is correlated with the relationship between
the respective jet energy scales. The likely size of any 
uncertainty in the measured jet masses is therefore assessed by rescaling
these in proportion to the corrections applied to the jet energies described 
above. Whereas the
true jet energies are known in Z calibration data and can therefore be 
corrected, the true jet masses are not. Consequently no corresponding 
corrections can be made for the jet mass scale, while the full 
size of the shift seen when rescaling the jet masses is applied as a 
systematic uncertainty. As the scale factors are determined from independent 
calibrations for each year, the effect of this is strongly year-dependent. 

The Z data are finally used to estimate the uncertainty in the
simulation of the electromagnetic calorimeter energy scale in hadronic events,
since the primary hadronic event selection relies on this. 

The largest influence on $\Delta\Ebeam$ arises from the uncertainty in the mass
scale of jets, with other notable contributions 
from uncertainties in the energy scales of jets and photons and the
angular scale of jets. Table~\ref{had_det} details these.  

\begin{table}[htb]
\begin{center}
\begin{tabular}{lcccc|c}
\hline
Detector effect & \multicolumn{5}{c}{Systematic Error /MeV} \\ \cline{2-6}
       & 1997 & 1998 & 1999 & 2000 & All years \\
\hline
Jet mass scale                     &$  8 $&$ 41 $&$ 13 $&$ 60 $&$ 25 $ \\
Jet energy scale                   &$ 16 $&$ 17 $&$ 18 $&$ 18 $&$ 17 $ \\
Photon energy scale                &$ 14 $&$ 13 $&$ 11 $&$  8 $&$ 12 $ \\
Jet angular scale                  &$  9 $&$  9 $&$  9 $&$  9 $&$  9 $ \\
Photon angular resolution          &$  2 $&$  3 $&$  5 $&$  7 $&$  4 $ \\
Photon energy linearity            &$  4 $&$  4 $&$  4 $&$  4 $&$  4 $ \\
Photon energy resolution           &$  2 $&$  3 $&$  4 $&$  6 $&$  3 $ \\
Jet energy resolution              &$  1 $&$  2 $&$  2 $&$  3 $&$  2 $ \\
Jet energy linearity               &$ <1 $&$  1 $&$  1 $&$  1 $&$  1 $ \\
ECAL energy scale                  &$  1 $&$  1 $&$  3 $&$  5 $&$  2 $ \\
Jet angular resolution             &$ <1 $&$ <1 $&$  1 $&$  1 $&$ <1 $ \\
\hline
{\bf Total} &$ {\bf 25} $&$ {\bf 47} $&$ {\bf 28} $&$ {\bf 65} $&$ {\bf 34}$ \\
\hline
\end{tabular}
\end{center}
\caption{\sl Detector modelling systematic error contributions on
 $\Delta\Ebeam$ for hadronic events.}
\label{had_det}
\end{table}

\item {\bf Fragmentation:} 
The sensitivity of the measurements to the fragmentation modelling of quarks
is estimated by comparing the \Pythia\ program (based on a
parton shower model and string hadronisation)
with \Herwig\  (parton shower model and cluster hadronisation) and 
\Ariadne\ (colour-dipole model and string hadronisation). In all cases the 
input parameters to the models are optimised through studies of global
event shape variables and particle production rates in calibration data taken
at the Z peak~\cite{bib:rates,bib:pr379}.  
To reduce statistical errors on this comparison, the same primary quarks 
generated with $\KK$ are fragmented according to each model in turn. The 
larger deviation from the \Pythia\ prediction arises from the comparison with
\Ariadne; the size of this deviation is assigned as a systematic error. 
The multiplicity cuts in the hadronic event selection are also varied by 
$\pm1$ unit to check the sensitivity to modelling of low multiplicity jets. 
This effect is negligible by contrast.  

\item {\bf Fit parameters:} 
The parameters fixed in the fits: $a$, $b$, $c$ and $\Gamma_{\pm}$, are 
varied by one standard deviation of their fitted values. For the first three 
of these, negligible shifts of the peak position, $M^*$, result. Although 
shifts of $M^*$ of up to $\sim 15$~MeV are observed in the
cases of the fitted widths, the change in Monte Carlo
is almost exactly mimicked by the corresponding change in the data. 
Accordingly a systematic uncertainty is assigned based on the residual
bias between Monte Carlo and data. 

\item {\bf ISR modelling:} 
The $\KK$ Monte Carlo is 
used as the default model for the $\mathrm{q\overline{q}}\Pgg$ process since 
it has the most complete available modelling of the ISR process, 
corresponding to $\mathcal{O}(\alpha^2)$ with CEEX.
The precision is degraded to correspond to $\mathcal{O}(\alpha)$ by a
reweighting procedure to give an estimate of the accuracy of the
description of ISR\@. Following
the recommendation of Ref.~\cite{bib:2fYR}, half of the difference observed
between the two schemes is assigned as a systematic error, reflecting the 
effects of missing higher order terms in the perturbative expansion.  
Further tests are performed against the Exclusive
Exponentiation (EEX) scheme~\cite{bib:eex} 
(the more primitive version of CEEX, formulated
in terms of spin-summed/averaged differential cross-sections rather than in 
terms of the more fundamental
spin amplitudes) at various orders. The results of all these checks, averaged 
over years, are detailed in Table~\ref{mcshifkk} for comparison.

\begin{table}[htb]
\begin{center}
\begin{tabular}{llccc}
\hline
\multicolumn{2}{c}{Scheme}  & 
\multicolumn{2}{c}{Shift in $\Delta\Ebeam$ /MeV} \\
\hline
$\KK$ weight & \multicolumn{1}{c}{Precision} & $\qq\gamma$ 
                                             & $\mumu\gamma$
                                             & $\tautau\gamma$ \\ 
\hline
CEEX2 (I/FSR interf.)
&   $\cal{O}${\small($1,\alpha,L\alpha,L^{2}\alpha^{2},
               L\alpha^{2}$)}              & --  & default & default \\
CEEX1 (I/FSR interf.) 
             &   $\cal{O}${\small($1,\alpha,L\alpha$)} & -- & $+1$ & $-13$ \\
CEEX0 (I/FSR interf.)
             &   $\cal{O}${\small($1$)}                & -- & $-11$& $+3 $ \\
\hline
CEEX2 (no I/FSR interf.)
&   $\cal{O}${\small($1,\alpha,L\alpha,L^{2}\alpha^{2},
               L\alpha^{2}$)}             & default  &$-4$ & $-4$ \\
CEEX1 (no I/FSR interf.)
             &   $\cal{O}${\small($1,\alpha,L\alpha$)}   
                                          & $-7$ &  $-2$ & $-20$ \\
\hline
EEX3 (no I/FSR interf.)  &   $\cal{O}${\small($1,\alpha,L\alpha,L^{2}
                          \alpha^{2},L^{3}\alpha^{3}$)} 
                                  & $-14$ & $  0$ & $-5 $\\
EEX2 (no I/FSR interf.) &   $\cal{O}${\small($1,\alpha,L\alpha,
                                     L^{2}\alpha^{2}$)} 
                                  & $-13$ & $ 0 $& $-5 $\\
\hline
\end{tabular}
\end{center}
\caption{\sl Shifts in $\Delta\Ebeam$, averaged over years,
due to different treatments of ISR\@. For the hadronic channel, a systematic
uncertainty is assigned as half of the difference between the CEEX2 (no I/FSR
interf.) and CEEX1 (no I/FSR interf.) schemes; for the leptonic channels,
half of the difference between the CEEX2 (I/FSR interf.) and CEEX1 (I/FSR 
interf.) schemes is taken.}
\label{mcshifkk}
\end{table}

\item {\bf Backgrounds:} 
The uncertainty arising from the estimation of the four-fermion background 
is assessed by comparing samples generated using \grcff\ and \Koralw. 
The difference between the two predictions has a negligible effect, as 
expected, since the largest component of 
this background, the $\PZ\Pep\Pem$ final state, can be regarded as signal-like.
The uncertainty from the untagged two-photon background is assessed by
comparing samples generated using \Phojet\ and \Pythia, from the tagged
two-photon background by comparing a combination of samples generated by
\Herwig\ and \Phojet\ with samples generated by \Twogen\@, and from the
$\tautau$ background by comparing samples generated using $\KK$ and 
\Koralz\@. These differences in prediction also have a negligible effect.

\item {\bf I/FSR interference:}
As explained in Section~\ref{sect-datamc}, the
Monte Carlo for hadronic events does not include the interference between
initial- and final-state photon radiation (I/FSR interference) which is 
naturally present in the data. To estimate the error introduced by the absence
of this effect in Monte Carlo, alternative samples of events were generated 
with FSR and I/FSR interference 
turned {\em on}\/ in the generation of the primary quark pairs in $\KK$, 
and FSR 
turned {\em off}\/ in their subsequent fragmentation, performed by 
\Pythia\@. A reweighting procedure enables these events
to be compared with the corresponding events should I/FSR interference have
instead been absent. Though FSR is incorrectly treated in this manner, the
effect cancels to some extent in comparing the weighted and unweighted events.
In any case, the negligible difference observed indicates that this concern 
is not a problem.

\item {\bf Beam energy spread/boost:} 
The effect of the finite spread of 
energies in the beams is to provide an event-by-event boost to the events,
corresponding typically to an rms spread of 250~MeV in the centre-of-mass
energy. In addition, there is a small net boost, with a typical
value of $-24$~MeV at the OPAL interaction point~\cite{lepewg},
caused by
asymmetries in the \LEP\ radio frequency accelerating system.
The size and spread of 
this boost is consistent with the muon-pair data (see
Section~\ref{sect-syst_l} below). The consequence of the first effect is 
investigated
by applying a Gaussian-distributed boost with mean zero and rms 250~MeV to the
Monte Carlo events, and of the second by applying a net boost of $-$24~MeV\@.
The sign indicates that the boost is in the $-z$ direction in \Opal.
The combined effect on $\Delta\Ebeam$ is found to be no more than 1~MeV\@.

\item {\bf Monte Carlo statistics:} 
The uncertainty resulting from limited 
Monte Carlo statistics is regarded as a systematic error, but is quoted 
separately.  

\item {\bf LEP calibration:} 
The error in the standard LEP determination of the beam energy~\cite{lepewg} 
contributes to the uncertainty in the difference between this and the value 
determined from OPAL data. Being unassociated with the details of our method,
it is quoted separately and is different in each year.

\end{itemize}
As a cross-check on the $\sqrt{s'}$ evaluation procedure,  
two alternatives are adopted. First, a simpler algorithm is used in which 
{\em exactly}
one ISR photon, either in the calorimeter or along the $z$-axis,
is allowed for {\em all} events. Second, an alternative set of cuts to
identify photons in the calorimeter is applied to the default
algorithm. Both give results consistent with the default;
no further error is therefore assigned.

\begin{table}[htb]
\begin{center}
\begin{tabular}{lcccc|c}
\hline
Effect & \multicolumn{5}{c}{Systematic Error /MeV} \\ \cline{2-6}
       & 1997 & 1998 & 1999 & 2000 & All years\\
\hline
Detector Modelling                 &$ 25 $&$ 47 $&$ 28 $&$ 65 $&$ 34 $ \\
Fragmentation                      &$ 13 $&$ 15 $&$ 18 $&$ 21 $&$ 16 $ \\
Fit parameters                     &$  4 $&$  1 $&$  5 $&$  4 $&$  3 $ \\
ISR modelling                      &$  3 $&$  3 $&$  3 $&$  4 $&$  3 $ \\
Backgrounds                        &$  1 $&$  1 $&$  1 $&$  2 $&$  1 $ \\
I/FSR interference                 &$  2 $&$  1 $&$  1 $&$ <1 $&$  1 $ \\
Beam energy spread/boost           &$  1 $&$  1 $&$ <1 $&$  1 $&$  1 $ \\
\hline
{\bf Total} &$ {\bf 29} $&$ {\bf 50} $&$ {\bf 33} $&$ {\bf 69} $&$ {\bf 38} $ \\
\hline
Monte Carlo statistics             &$ 12 $&$ 10 $&$  7 $&$  7 $&$  5 $ \\
LEP calibration                    &$ 10 $&$ 11 $&$ 12 $&$ 20 $&$ 11 $ \\
\hline
{\bf Full Total} &$ {\bf 33} $&$ {\bf 52} $&$ {\bf 36} $&$ {\bf 72} $&$ {\bf 40} $ \\
\hline
\end{tabular}
\end{center}
\caption{\sl Systematic error contributions
on $\Delta\Ebeam$ for hadronic events.}
\label{had_syst}
\end{table}


\subsection{Leptonic Channels}
\label{sect-syst_l}

The following effects are taken into account, and the uncertainties are 
summarised in Table~\ref{syst-lep}.
\begin{itemize}

\item {\bf Lepton angular scale:} 
The measurement is sensitive to any bias in the
the reconstructed direction of tracks, clusters or cones, in particular
the $\theta$ measurement (since the majority of events are
those with the photon along the beam direction.) 

The analysis for the muon events is repeated using the measured $\theta$
value of the associated electromagnetic energy cluster 
(shift of +24~MeV in $\Delta \Ebeam$)
or track segment in the muon chamber (shift of +41~MeV). 
These shifts are consistent with the 
rms shift estimated by 
an approximate Monte Carlo study, in which the 
track $\theta$ measurement is 
shifted and smeared according to the 
mean and rms of the differences seen
in data between the default track measurement and
the alternative calorimeter or muon chamber measurement.
The position of 
lead-glass blocks in the calorimeter is determined by the
known geometry and survey information, and is independent
of the tracking. There are known problems with modelling
the energy deposition and apparent angle of minimum
ionising particles especially
in the endcap lead-glass. The track measurement can 
therefore be considered more reliable.
The barrel muon chambers are partly calibrated against
tracks, while the information
from the end-cap muon chambers is more independent.
A systematic error of 21~MeV is assigned, equal
to half of the larger shift seen, i.e. resulting
from the comparison of tracking and muon chamber 
information.

The $\theta$ angle of the tau cone is reevaluated using
tracks only (shift of $+131$~MeV) or clusters only 
($-22$~MeV). A similar Monte Carlo study 
to that for the muon events confirmed that
the shifts are consistent with the statistical 
uncertainty associated with the degradation in 
precision expected from  
removing clusters or tracks from
the angle determination.
A systematic uncertainty of 66~MeV is assigned, equivalent
to half the larger shift.

Similarly, the $\theta$ angle of electron candidates is
replaced by the direction of the associated track (shift of
$-48$~MeV). There is a problem with the modelling of 
high energy, fairly forward electron tracks, since electrons tend
to radiate in the tracking volume, unlike muons.
Again, half the shift, 24~MeV, is assigned as the 
systematic uncertainty.
\item {\bf Lepton angular resolution:} The modelling of the
$\theta$ resolution is checked by examining the distribution of
$\cos \theta_1 + \cos \theta_2$ for full-energy, back-to-back
events. For muon and electron events, the resolution in data is worse
than in the Monte Carlo, while for tau events the Monte Carlo
resolution is slightly worse than that of the data. Part of the
disagreement could be accounted for by the spread in 
centre-of-mass energy described below.
The $z$-momentum in the Monte Carlo is smeared so as to 
bring the muon and electron distributions into agreement with 
the data, and by a similar amount in the tau-channel 
to estimate the systematic uncertainty.

\item {\bf Fit parameters:} The widths of the Breit-Wigner 
distribution are varied by their fitted errors, and the positions of the $M^*$ 
peaks in data and Monte Carlo redetermined.

\item {\bf ISR modelling:} To evaluate the sensitivity to the modelling
of ISR, the analysis is repeated, reweighting the $\KK$ Monte Carlo
samples to other schemes. Samples of muon and tau pairs with
event weights
for the different schemes
are available at 189~GeV and 206~GeV.
The CEEX scheme sometimes fails for
muon events, in which case the EEX3 scheme is used.
Very large weights are sometimes generated for tau events
with a low tau-pair mass; weights larger than 10.0 are
taken to be equal to 10.0.
As can be seen in Table~\ref{mcshifkk},
the tau events show larger shifts than the muon sample.
Following the recommendations described above for the hadrons,
half the difference between the CEEX2 and CEEX1 models 
(with interference between initial-
and final-state radiation for leptons) is
taken as a systematic uncertainty, {\em i.e.}\/ $+1$~MeV and $-7$~MeV
for the muon  and tau events respectively. 
The BHWIDE Monte Carlo is used for the electron channel,
with calculations of order $\cal{O}${\small($\alpha$)}
with YFS exponentiation. Reweighting events to 
switch off the exponentiation,
a shift of $4\pm20$~MeV is observed, where the error is
statistical. An uncertainty of 10~MeV is assigned,
equal to half the precision of this test.

\item {\bf Backgrounds:} Varying the small background in the muon
sample has a negligible effect on the result. 
The two-photon, four-fermion and Bhabha backgrounds
in the tau sample are each varied by $\pm 10\%$. This range is 
motivated by the discrepancies in the number of events below and
above the $\Sigma E/ \sqrt{s} $ range accepted in the tau event selection.
The non-resonant background in the electron sample is
also varied by $\pm 10\%$, and the rate and slope of the fitted
$t$-channel contribution are shifted by the fitted errors.

\item {\bf Beam energy spread/boost:}
The mean and width of the distribution of $\cos \theta_1 + \cos \theta_2$
for non-radiative Monte Carlo simulated muon-pair events is
in reasonable agreement with the data when 
an average boost of
$-24$~MeV with an rms spread of 250~MeV is applied
to the simulation. The changes in $\Delta\Ebeam$
from applying these boosts to the simulation
are assigned as a systematic uncertainty.

\item {\bf Monte Carlo statistics:}
The uncertainty resulting from limited 
Monte Carlo statistics is regarded as a systematic error, but is quoted 
separately.  
\item {\bf LEP calibration:} 
The error in the standard LEP determination of the beam energy~\cite{lepewg} 
contributes to the uncertainty in the difference between this and the value 
determined from OPAL data. Being unassociated with the details of our method,
it is quoted separately, averaged over years.

\end{itemize}
Tests with low statistics Monte Carlo samples give no indication
of a bias in the method, and suggest that the errors from the
fits are reasonable.

\begin{table}[htb]
\begin{center}
\begin{tabular}{lccc}
\hline
Effect & \multicolumn{3}{c}{Systematic Error /MeV} \\ \cline{2-4}
       & $\mumu\gamma$ & $\tautau\gamma$ & $\epem\gamma$ \\
\hline
Lepton angular scale         &$ 21 $&$  66  $&$  24 $\\
Lepton angular resolution    &$  2 $&$   4  $&$   7 $\\ 
Fit parameters               &$  1 $&$   4  $&$  10 $\\
Non-resonant background      &$ <1 $&$   6  $&$   4 $\\
Bhabha/$t$-channel           &$ <1 $&$   3  $&$   5 $\\
ISR modelling                &$  1 $&$   7  $&$  10 $\\ 
Beam energy spread/boost     &$  2 $&$   5  $&$   6 $\\ 
\hline
{\bf Total}          & {\bf 21} & {\bf  67}& {\bf 30}\\ 
\hline
Monte Carlo statistics       & $9  $&$ 34  $&$ 34   $\\
LEP calibration              & $11 $&$ 11  $&$ 11   $\\
\hline
{\bf Full Total} & {\bf 25} & {\bf  76}& {\bf 46 }\\ 
\hline
\end{tabular}
\end{center}
\caption{\sl Systematic error contributions
on $\Delta\Ebeam$ for leptonic events.}
\label{syst-lep}
\end{table}


\section{Discussion and Summary}
\label{sect-summ}

\begin{figure}
  \epsfxsize=\textwidth
  \epsffile{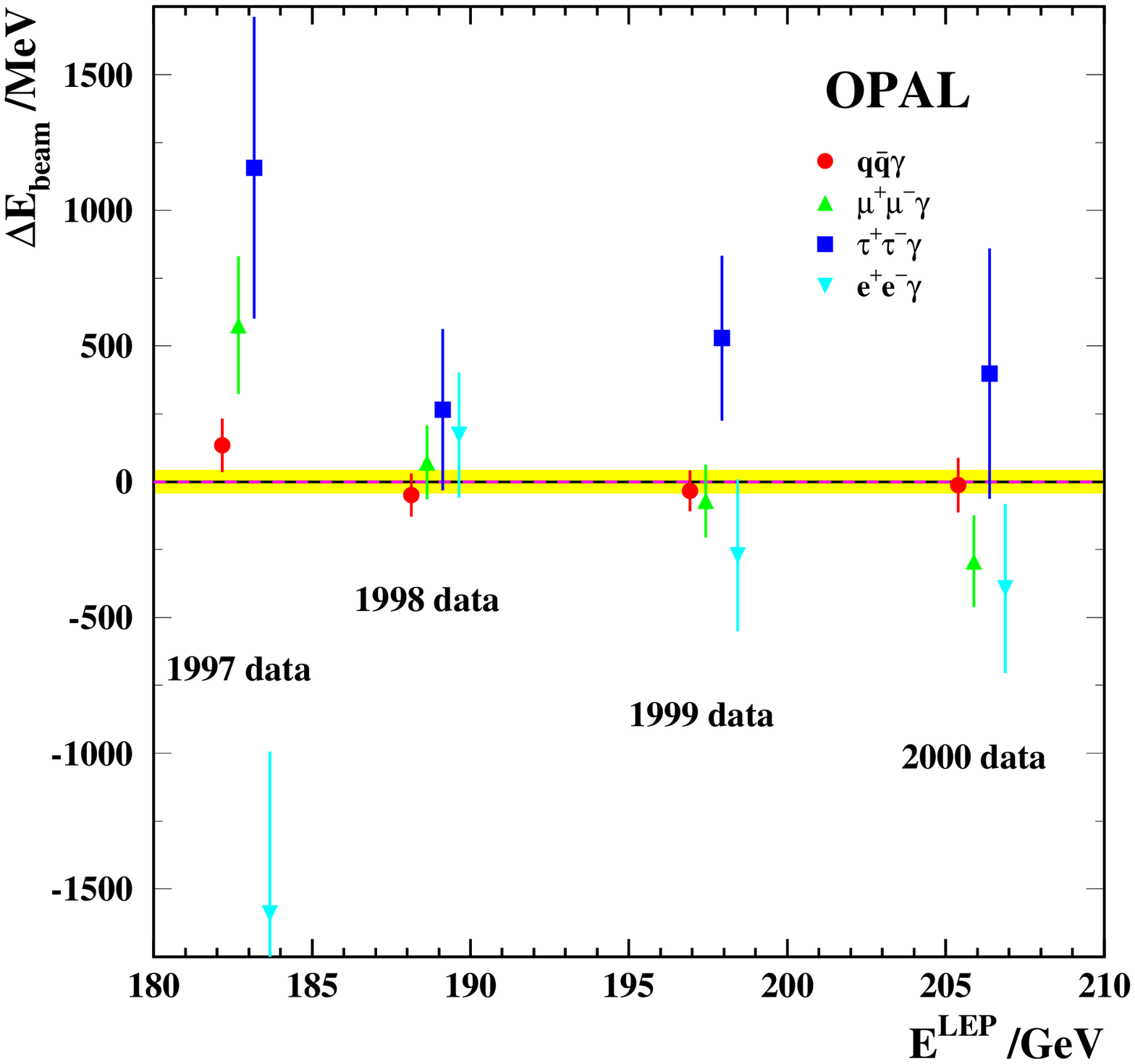}
  \caption{\sl Summary of measured values of 
$\Delta\Ebeam$, using hadronic and leptonic events in \Opal\ data, as a 
function of the centre-of-mass energy. 
For clarity, measurements made with hadrons have been displaced leftwards by
0.5~GeV, while those made with tau and electron pairs have been displaced
rightwards by 0.5~GeV and 1.0~GeV respectively. The dashed line represents 
the overall average, with the shaded band indicating its total error, including the 11~MeV uncertainty from the standard \LEP\ beam energy determination.}
\label{fig:lepecmplot}
\end{figure}

Using fermion-pair events at \Leptwo\ which exhibit radiative return to 
the Z, together with knowledge of the Z mass, we have made estimates of the 
\LEP\ beam energy using \Opal\ data.   
In Fig.~\ref{fig:lepecmplot}  we show a summary of the measurements
of  $\Delta\Ebeam$ using hadronic and leptonic
final states at centre-of-mass energies 
from 183~GeV to 209~GeV\@. 
There is no significant evidence for any dependence on
centre-of-mass energy. Average values for each channel, 
and for each year of data-taking, are summarised in 
Table~\ref{tab:results-all}.   Common systematic uncertainties are
taken into account in forming the averages. 
For example, detector systematics
for the hadron results are taken to be fully correlated from year to year, as
are fragmentation systematics. However, detector systematics are assumed not 
to be correlated between hadrons and leptons, nor between different lepton 
species.
The combined value for all energies from hadronic events is
\[ \Delta E_{\mathrm{b}} = +1\pm 38(\mathrm{stat.})\pm 40(\mathrm{syst.})
\;\mathrm{MeV},\]
while from leptonic events this is
\[ \Delta E_{\mathrm{b}} = -2\pm 62 (\mathrm{stat.})\pm 24(\mathrm{syst.})
\;\mathrm{MeV}.\] 
Both are evidently consistent with zero.

If all the results are combined, weighting the measurements using the
total errors, and assuming the systematic errors to be 
uncorrelated between the hadronic and leptonic channels except for those
associated with \ISR\ modelling, the beam energy spread and boost and, in part,
the LEP calibration, the overall estimate of the shift in the beam energy is
\[ \Delta E_{\mathrm{b}} = 0 \pm 34 (\mathrm{stat.})\pm 27 (\mathrm{syst.})
\;\mathrm{MeV}.\]
The uncertainty from the standard LEP beam energy determination contributes
11~MeV to the systematic error. 

We therefore see no evidence of any disagreement between the
\Opal\ data and the standard \LEP\ energy calibration, either overall or in
any year of data-taking. Combination with similar results from other \LEP\ 
experiments~\cite{bib:others} should allow a more precise comparison with the beam energy 
determined by \LEP\@.

\section*{Acknowledgments}

We particularly wish to thank the SL Division for the efficient operation
of the LEP accelerator at all energies
 and for their close cooperation with
our experimental group.  In addition to the support staff at our own
institutions we are pleased to acknowledge the  \\
Department of Energy, USA, \\
National Science Foundation, USA, \\
Particle Physics and Astronomy Research Council, UK, \\
Natural Sciences and Engineering Research Council, Canada, \\
Israel Science Foundation, administered by the Israel
Academy of Science and Humanities, \\
Benoziyo Center for High Energy Physics,\\
Japanese Ministry of Education, Culture, Sports, Science and
Technology (MEXT) and a grant under the MEXT International
Science Research Program,\\
Japanese Society for the Promotion of Science (JSPS),\\
German Israeli Bi-national Science Foundation (GIF), \\
Bundesministerium f\"ur Bildung und Forschung, Germany, \\
National Research Council of Canada, \\
Hungarian Foundation for Scientific Research, OTKA T-038240,
and T-042864,\\
The NWO/NATO Fund for Scientific Research, the Netherlands.\\


\normalsize

\begin{thebibliography}{99}

\bibitem{yellowbook}
A.~Ballestrero et al., ``Physics at LEP2'' p.141; CERN 96-01 vol 1.  

\bibitem{lepewg}
\LEP\ Energy Working Group, A.~Blondel et al., Eur.~Phys.~J.~{\bf C11} 
      (1999) 573;\\ 
\LEP\ Energy Working Group, R.~Assmann et al., CERN-PH-EP/2004-032,
submitted to Eur.~Phys.~J.

\bibitem{pdg}
Particle Data Group, K.~Hagiwara et al., Phys.~Rev.~{\bf D66} (2002) 010001.

%



\bibitem{bib:OPAL-detector}
  \OPALColl, K.~Ahmet et~al., \NIM\ {\bf A305} (1991) 275.

\bibitem{bib:OPAL-SI}
  S.~Anderson et~al., \NIM\ {\bf A403} (1998) 326.

\bibitem{bib:OPAL-TR}
  M.~Arignon et~al., \NIM\ {\bf A313} (1992) 103; \\
  M.~Arignon et~al., \NIM\ {\bf A333} (1993) 330.

\bibitem{bib:OPAL-DAQ}
  J.T.M.~Baines et~al., \NIM\ {\bf A325} (1993) 271; \\
  D.G.~Charlton, F.~Meijers, T.J.~Smith and P.S.~Wells, \NIM\ {\bf A325} 
  (1993) 129.

\bibitem{bib:OPAL-SW}
  OPAL Collab., G.~Abbiendi et~al., 
  \EPJ\ {\bf C14} (2000) 373.

\bibitem{bib:GOPAL}
  \OPALColl, J.~Allison et~al., \NIM\ {\bf A317} (1992) 47.

\bibitem{bib:KK2f}
  S.~Jadach, B.F.L.~Ward and Z.~W\c{a}s, \CPC\ {\bf 130} 
  (2000) 260.

\bibitem{bib:ceex}
  S.~Jadach, B.F.L.~Ward and Z.~W\c{a}s, \PhysRev\ {\bf D63} (2001) 113009.

\bibitem{bib:bhwide}
  S.~Jadach, W.~Placzek and B.F.L.~Ward, \PhysLett\ {\bf B390} (1997) 298.

\bibitem{bib:yfs} 
D.R.~Yennie, S.C.~Frautschi and H.~Suura. Ann.~Phys.~(N.Y.) {
\bf 13} (1961) 379.

\bibitem{bib:PYTHIA}
  T.~Sj\"ostrand et~al., \CPC\ {\bf 135} (2001) 238.


\bibitem{bib:HERWIG}
  G.~Corcella et al.,  J.~High Energy Phys.~{\bf 0101} (2001) 010.

\bibitem{bib:ARIADNE}
  L.~L\"onnblad, \CPC\ {\bf 71} (1992) 15.

\bibitem{bib:GRC4F}
J.\ Fujimoto et al., \CPC\ {\bf 100} (1997) 128.

\bibitem{bib:KORALW}
S.~Jadach et~al.,  \CPC\ {\bf 119} (1999) 272.

\bibitem{bib:phojet}
  R.~Engel and J.~Ranft, \PhysRev\ {\bf D54} (1996) 4244.

\bibitem{bib:twogen}
  A.~Buijs et al., \CPC\ {\bf 79} (1994) 523.

\bibitem{bib:verm}
J.A.M.~Vermaseren, Nucl.~Phys.~{\bf B229} (1983) 347.

\bibitem{bib:KORALZ}
  S.~Jadach, B.F.L.~Ward and Z.~W\c{a}s, \CPC\ {\bf 79} (1994) 503.

\bibitem{bib:cga_thesis}
C.G.~Ainsley, {\em Studies of $Z/\gamma\rightarrow q\overline{q}$ 
events with the OPAL detector at LEP~II}, Ph.D. Thesis, 
University of Cambridge, RAL-TH-2003-001, January 2003.

\bibitem{bib:OPAL-SM209}
  OPAL Collab., G.~Abbiendi et~al., 
  \EPJ\ {\bf C33} (2004) 173.

\bibitem{bib:OPAL-SM189}
  OPAL Collab., G.~Abbiendi et~al., 
  \EPJ\ {\bf C13} (2000) 553.

\bibitem{bib:OPAL-SM183}
  OPAL Collab., G.~Abbiendi et~al., 
  \EPJ\ {\bf C6} (1999) 1.

\bibitem{bib:OPAL-SM172}
  OPAL Collab., K.~Ackerstaff et~al., 
  \EPJ\ {\bf C2} (1998) 441.

\bibitem{Durham}
S.~Catani et al., Phys.~Lett.~{\bf B269} (1991) 432.

\bibitem{bib:MT}
  OPAL Collab., G.~Abbiendi et~al.,
  \EPJ\ {\bf C26} (2003) 479.

\bibitem{bib:OPAL-MW189}
  OPAL Collab., G.~Abbiendi et~al., \PhysLett\ {\bf B507} (2001) 29.



\bibitem{bib:rates}
  OPAL Collab., G.~Alexander et~al., \ZPhys\ {\bf C69} (1996) 543.

\bibitem{bib:pr379}
  OPAL Collab., G.~Abbiendi et~al., 
  \EPJ\ {\bf C35} (2004) 293.

\bibitem{bib:2fYR}
  Reports of the Working Groups on Precision Calculations for LEP2 Physics:
  Two-fermion Production in Electron-Positron Collisions, CERN-YR-2000-009, 
  hep-ph/0007180.

\bibitem{bib:eex}
  S.~Jadach and B.F.L.~Ward, \CPC\ {\bf 56} (1990) 351.

\bibitem{bib:others}
ALEPH Collab., R.~Barate et~al., \PhysLett\ {\bf B464} (1999) 339;\\
L3 Collab., P.~Achard et~al., \PhysLett\ {\bf B585} (2004) 42.





\end{thebibliography}
\end{document}